\documentclass{article}
\usepackage{spconf,amsmath,graphicx}
\usepackage{enumitem}
\setlist{nosep, leftmargin=14pt}

\title{Graph neural network for cerebral blood flow prediction with clinical datasets}

\name{Seungyeon Kim$^{a,b,1}$ \qquad Wheesung Lee$^{a,1}$ \qquad Sung-Ho Ahn$^{c,d}$ \qquad Do-Eun Lee$^{d}$ \qquad Tae-Rin Lee$^{a,\dagger}$}

\address{
$^{a}$R\&D Center, NEAR Brain Inc., Seoul, Republic of Korea\\
$^{b}$Department of Applied Artificial Intelligence, Sungkyunkwan University, Suwon, Republic of Korea\\
$^{c}$Research Institute for Convergence of Biomedical Science and Technology, \\ Pusan National University School of Medicine, Busan, Republic of Korea\\
$^{d}$Department of Neurology, Pusan National University Yangsan Hospital, Busan, Republic of Korea\\
$^{1}$These authors are equally contributed.\\
$^{\dagger}$Corresponding author, taerin@nearbrain.com}

\begin{document}
\maketitle

\begin{abstract}
\noindent Accurate prediction of cerebral blood flow is essential for the diagnosis and treatment of cerebrovascular diseases. Traditional computational methods, however, often incur significant computational costs, limiting their practicality in real-time clinical applications. This paper proposes a graph neural network (GNN) to predict blood flow and pressure in previously unseen cerebral vascular network structures that were not included in training data. The GNN was developed using clinical datasets from patients with stenosis, featuring complex and abnormal vascular geometries. Additionally, the GNN model was trained on data incorporating a wide range of inflow conditions, vessel topologies, and network connectivities to enhance its generalization capability. The approach achieved Pearson's correlation coefficients of 0.727 for pressure and 0.824 for flow rate, with sufficient training data. These findings demonstrate the potential of the GNN for real-time cerebrovascular diagnostics, particularly in handling intricate and pathological vascular networks.
\end{abstract}


\section{Introduction}
Various simulation models have been developed to predict cerebral blood flow (CBF) in the brain. For example, Moore et al.~\cite{moore2005one} developed a 1D flow model for the circle of Willis (CoW) and compared the simulation results with 3D computational fluid dynamics simulation. Alastruey et al.~\cite{Alastruey2007modelling} assessed the effect of CoW geometry variations on cerebral flow. In 2009, Grinberg et al.~\cite{Grinberg2009simulation} simulated the human intracranial arterial tree using an unsteady 3D flow model with one billion degrees of freedom. Perdikaris et al.~\cite{perdikaris2016multiscale} proposed a multiscale model for brain blood flow. Pegolotti et al.~\cite{pegolotti2024learning} suggested a reduced-order model for cardiovascular system based on graph neural network (GNN). Nevertheless, current simulation models are complex and based on 3D vascular structures, which would be challenging in clinical studies. 

In this study, we propose a development process of GNN to predict blood flow in the cerebrovascular system. Specifically, the patient group is focused on stenosis cases on the middle cranial artery (MCA). The reference datasets of patients are based on MR angiography-Time of Flight (MRA-TOF). A mathematical model for blood flow prediction is engaged to amplify training and testing datasets for GNN. The datasets for GNN are consist of various flow input conditions, geometry and connectivity to test the performance. 

\section{Method}
\subsection{Vessel Segmentation and Refinement}
Segmentation of blood vessels is performed after importing magnetic resonance angiography (MRA) images in NIfTI format. The images are preprocessed by smoothing with Gaussian filters, followed by hysteresis thresholding to remove noise and detect vessel voxels~\cite{van2014scikit}. The coordinates of points above the threshold are extracted to form a point cloud. This point cloud data is further refined using DBSCAN~\cite{pedregosa2011scikit} clustering to remove noise and filter out non-vessel points. Based on these results, the indices of points identified as vessel voxels are set to 1, reconstructing a 3D array.

\begin{figure*}[htb]
  \centering
  \includegraphics[width=\textwidth]{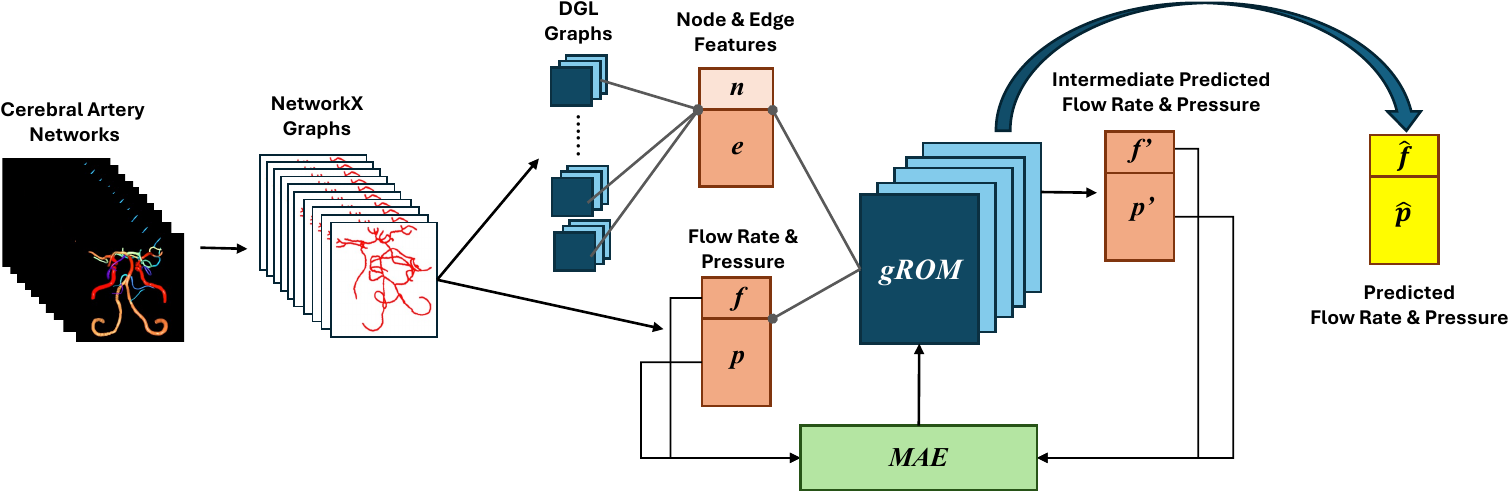}  
  \caption{Overall architecture of the proposed gROM framework for predicting flow rate and pressure in cerebral artery networks. The process begins with cerebral artery network data, which is transformed into NetworkX graphs. These graphs are further converted into DGL graphs containing node and edge features. The gROM takes these features along with flow rate (\( f \)) and pressure (\( p \)) inputs and applies the gROM for prediction. The intermediate predicted flow rate (\( f' \)) and pressure (\( p' \)) are first computed, and after further refinement, the final predicted flow rate (\( \hat{f} \)) and pressure (\( \hat{p} \)) are obtained. These final predicted values are compared to the ground truth using the Mean Absolute Error (MAE) as the loss function.}
  \label{fig:overall_architecture}
\end{figure*}

\subsection{Link Prediction and Network Construction}
To create a NetworkX graph~\cite{hagberg2008exploring}, each voxel in the array is examined to determine its neighbors, and the voxels are classified as either branch nodes or non-branch nodes. As the neighbors of each voxel are checked, edges are created by connecting to branch node voxels that share a cluster number. The coordinates of the nodes and edges are then stored, and the radius of each node is calculated. The final network stores each node's ID, position, and radius, as well as each edge's ID, position, length, and axis information.

\subsection{Mathematical Model for Blood Flow Prediction}
The movement of blood through a vessel segment can be assumed as a Poiseuille flow model:  

\begin{equation}
Q = \frac{\pi d^4}{128 \mu L}\Delta P
\label{eq:poisuille}
\end{equation}

\noindent where $Q$ is the flow rate in the blood vessel, $d$ is the vessel diameter, $L$ is the vessel length, $\Delta P$ is the pressure drop between the i-th and j-th nodes and $\mu$ is the blood viscosity. In addition, if the two diameters at the i-th and j-th nodes ($d_i$ and $d_j$) are different, the average value, \textit{i.e.} $d = (d_i+d_j)/2$, is used to calculate Eq.~\eqref{eq:poisuille}. The pressure values at the nodes of the vascular geometry were the unknown we wanted to solve. To solve the unknown pressure values, the conservation law of flow rates at a bifurcation can be defined as: 

\begin{equation}
\Sigma Q_i = 0
\label{eq:sumofq}
\end{equation}

\noindent where $Q_i$ is the flow rates at the i-th bifurcation junction. More details are given in \cite{choquantifying}.

\subsection{GNN Model}
\textbf{MeshGraphNets} MeshGraphNets~\cite {pfaff2020learning} is structured with an Encoder, Processor, and Decoder. The Encoder converts the physical properties of nodes and geometric information of edges into latent representations. In the Processor, nodes exchange information through a message-passing mechanism across edges, allowing physical and geometric data to propagate throughout the mesh. Finally, the Decoder infers changes in the physical quantities from the latent node vectors, predicting the system's dynamics for the next time step. 

\noindent\textbf{gROM} The gROM (Reduced-Order Models with GNNs)~\cite{pegolotti2024learning} is inspired by MeshGraphNets, adapting its core principles for real-time simulations of blood flow in cardiovascular systems. While MeshGraphNets are designed to simulate high-fidelity physical dynamics, gROM focuses on improving efficiency by simplifying the model for real-time simulations. 

\section{Experiments}
\subsection{Experimental Setup}

\noindent \textbf{Dataset Acquisition} In this study, we utilized MRA datasets, comprising a total of 35 cases. The dataset includes 25 patients diagnosed with cerebral infarction, 3 with transient ischemic attacks, 2 with basilar artery stenosis, and 5 with other cerebrovascular conditions, such as paresthesia, vascular dementia, and seizures. The patients ranged in age from 42 to 84 years (mean age: 65.8 years), with a gender distribution of 21 males and 14 females. MRA-TOF scans were performed using 3T MRI systems (Verio or Skyra, Siemens Healthineers, Germany). The imaging parameters for MRA-TOF included an echo time of 3.7 ms, repetition time of 21 ms, flip angle of 18°, a field of view of 167 × 260 mm, slice thickness of 0.5 mm, and an inter-slice spacing of 20 mm.

\noindent \textbf{Dataset Preprocessing} The original dataset of 35 individuals was processed into graph structures containing information on graph directionality, multigraph status, node attributes, edge attributes, inlet node IDs, and outlet node IDs. These graph representations were then augmented to create 25 datasets per individual, resulting in a total of 875 datasets. During the augmentation process, inlet pressure was randomly set between 12,000 Pa and 18,000 Pa, and the vessel radius was randomly adjusted between 0.8 and 1.2 times its original size. The augmentation aimed to enhance the model’s generalizability by introducing variability within the dataset. Given that the dataset included stenosis cases, the adjustments in vessel radius were made to reflect realistic pathological ranges~\cite{gharahi2016computational}. 

\noindent \textbf{Implementation Details} For training the GNN, the initial learning rate was set to 0.001, with a gradual decrease following a cosine annealing schedule. The minimum learning rate was set to $10^{-6}$, and training was conducted over a total of 500 epochs. The Mean Absolute Error (MAE) was used as the loss function, and the Adam optimizer was employed to optimize the model. 

\noindent \textbf{Evaluation Methods} To ensure a robust evaluation of the GNN's performance, 5-fold cross-validation was employed. During each fold, the GNN was trained using data from 28 distinct network structures and tested on 7 completely different network structures, with each network expanded to 25 datasets. The version selected for evaluation in each fold was the one that achieved the best performance, as measured every 100 epochs. Performance evaluation was conducted by comparing the GNN’s predictions of blood flow and pressure values with the results generated by the mathematical model.

The accuracy (\( R \)) was calculated based on the difference between the predicted values (both blood flow and pressure) from gROM and the actual values derived from the mathematical model. Specifically, accuracy was measured by counting the number of nodes where the error margin between the predicted and actual values was within 10\%. The formula used to compute this accuracy is given as follow:

\begin{equation}
R = \frac{1}{n} \sum \left( H \left( \frac{|v_{\text{pred}} - v_{\text{true}}|}{\max(v_{\text{true}})} \right) \right) \times 100(\%)
\end{equation}

\noindent In this equation, \( v_{\text{pred}} \) represents the predicted values, while \( v_{\text{true}} \) refers to the actual values obtained from the cerebral vascular model. The term \( \max(v_{\text{true}}) \) is the maximum value of \( v_{\text{true}} \), used to normalize the error. The function \( H(x) \), defined as:

\begin{equation}
H(x) = \begin{cases} 
1, & \text{if } x < 0.1 \\
0, & \text{otherwise}
\end{cases}
\end{equation}

\noindent returns 1 if the normalized error between the predicted and actual values is less than 10\%, and 0 otherwise. The overall accuracy (\( R \)) is the percentage of nodes where the error falls within this threshold.

Additionally, to analyze the correlation between the predicted and actual values of flow rate and pressure, the Pearson's correlation coefficient was employed as a metric for accuracy evaluation.

\begin{figure}[htb]
\centering

\begin{minipage}[b]{.48\linewidth}
  \centering
  \centerline{\includegraphics[width=\linewidth]{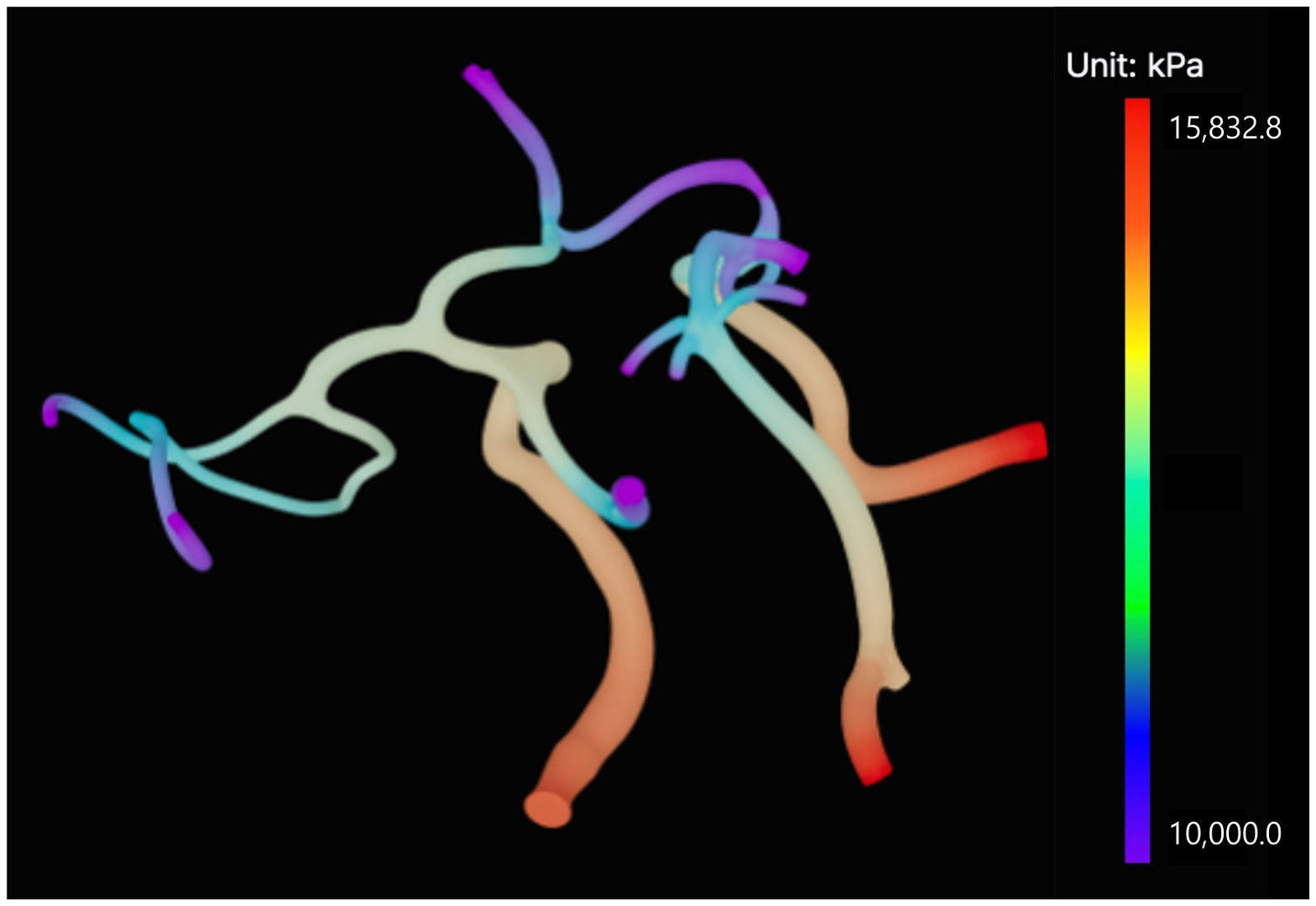}} 
  \centerline{(a) Ground truth pressure}
\end{minipage}
\hfill
\begin{minipage}[b]{.48\linewidth}
  \centering
  \centerline{\includegraphics[width=\linewidth]{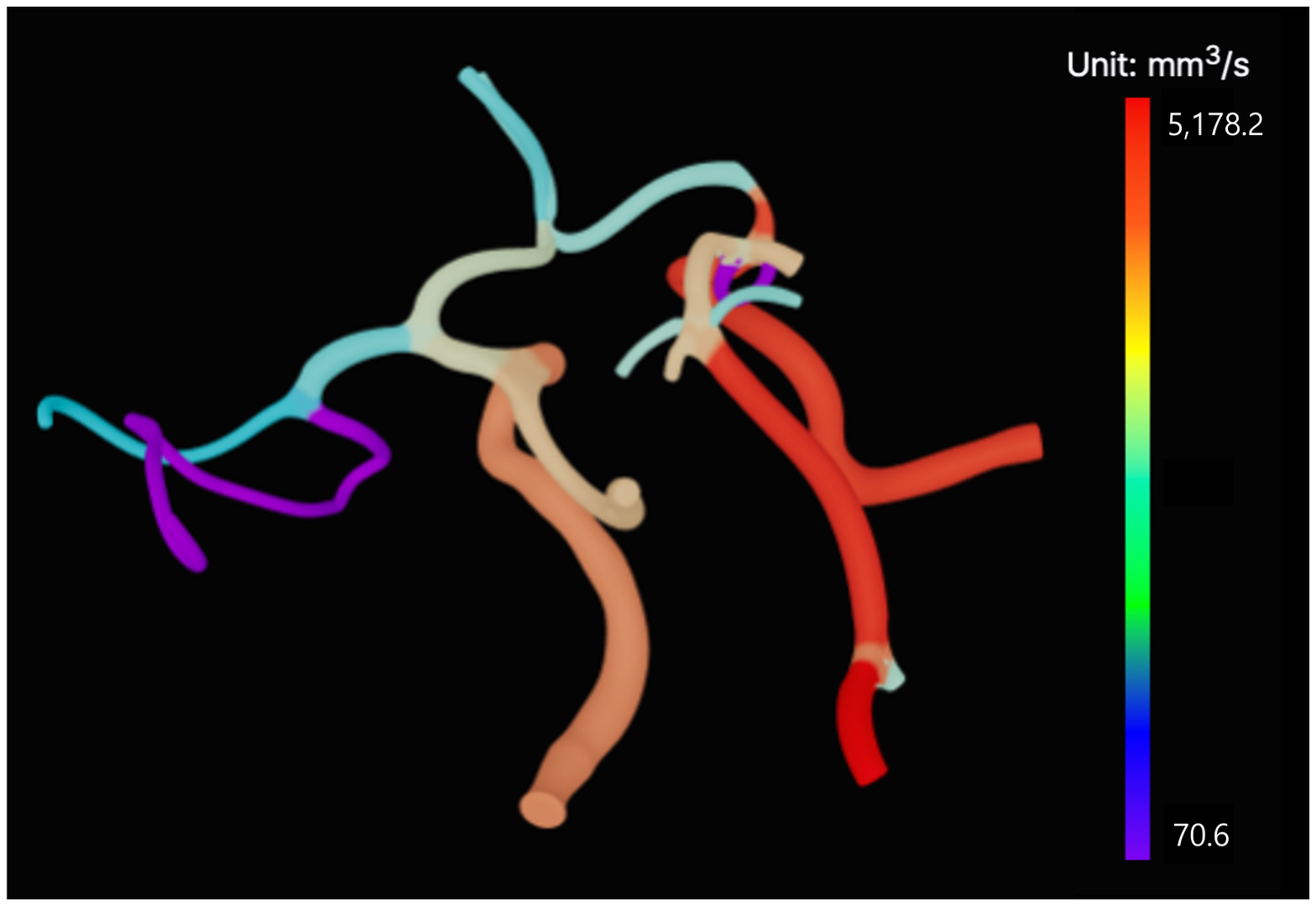}}
  \centerline{(b) Ground truth flow rate}
\end{minipage}
\vspace{0.5cm}  

\begin{minipage}[b]{.48\linewidth}
  \centering
  \centerline{\includegraphics[width=\linewidth]{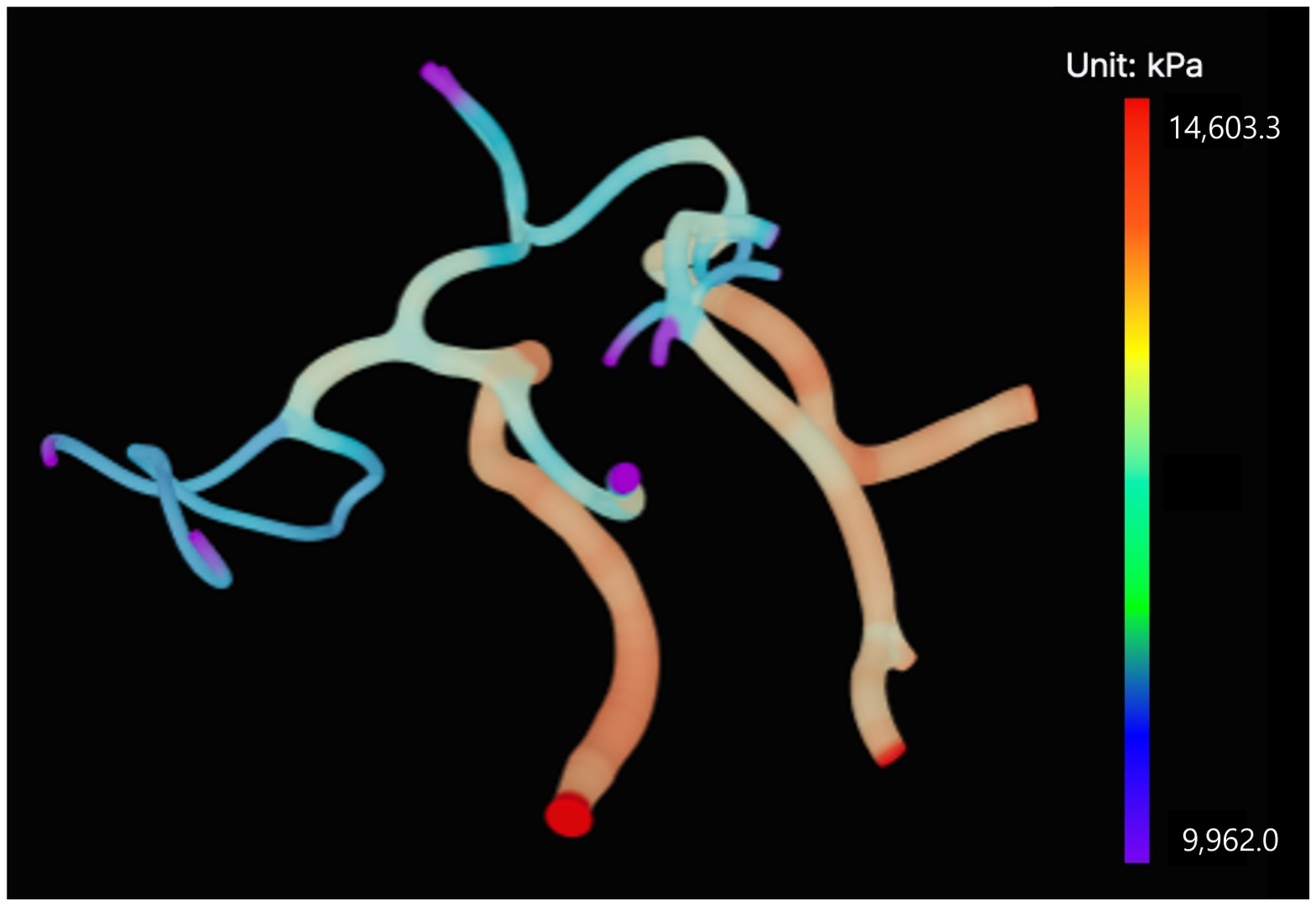}} 
  \centerline{(c) Predicted pressure}
\end{minipage}
\hfill
\begin{minipage}[b]{.48\linewidth}
  \centering
  \centerline{\includegraphics[width=\linewidth]{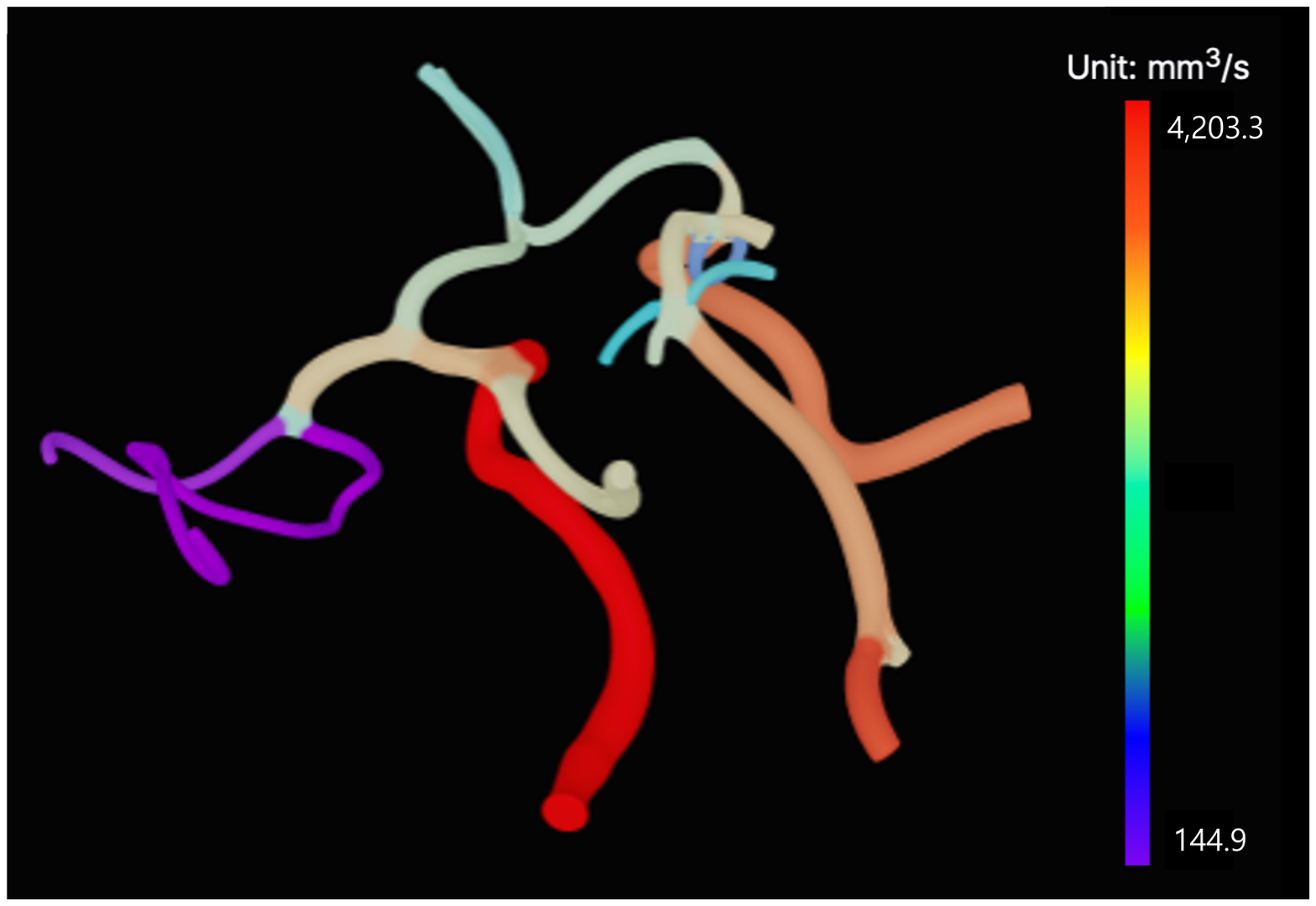}} 
  \centerline{(d) Predicted flow rate}
\end{minipage}

\caption{Visualization of pressure and flow rate using Dr. NEAR flow, comparing ground truth and model predictions.}
\label{fig:visualization_comparison}

\end{figure}

\begin{figure}[htb]
\centering
\begin{minipage}[b]{0.49\linewidth}
  \centering
  \includegraphics[width=\linewidth]{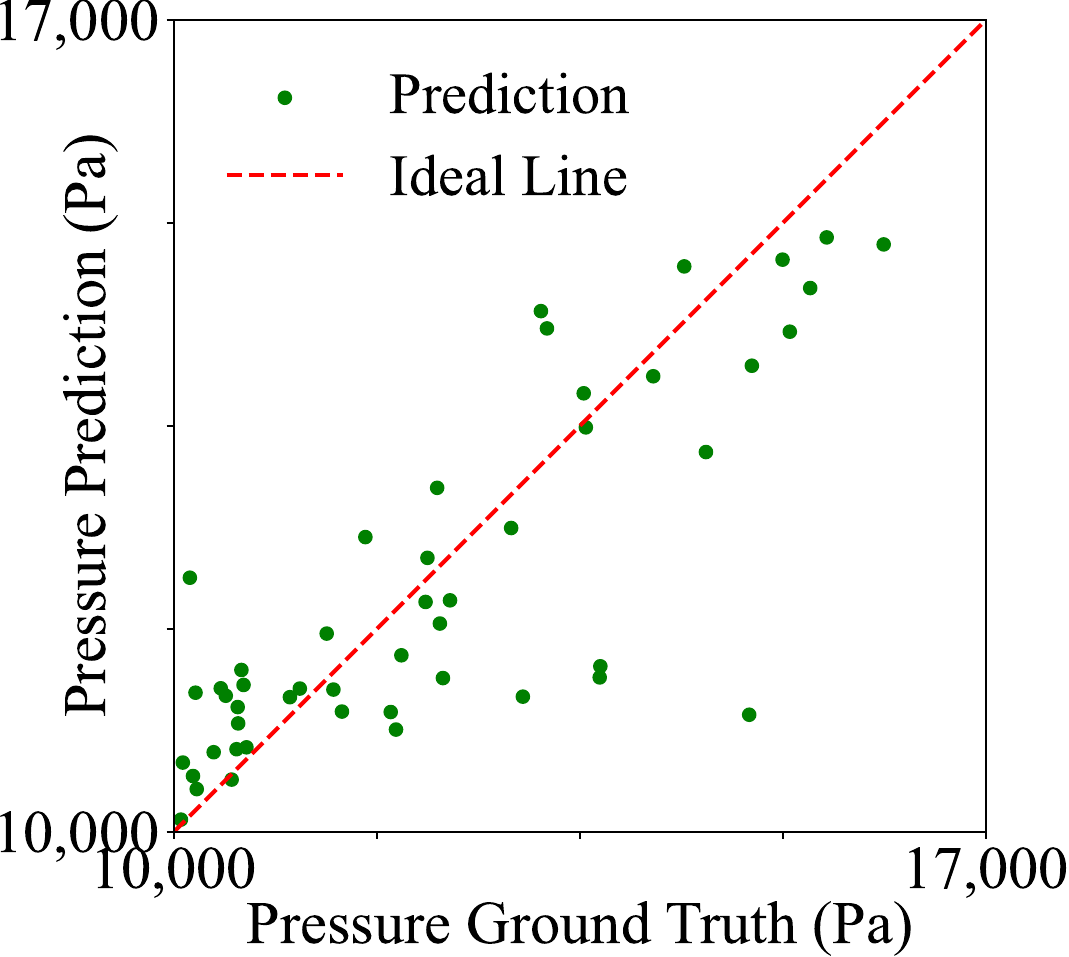}
  \centerline{(a) Pressure}
\end{minipage}
\hfill
\begin{minipage}[b]{0.49\linewidth}
  \centering
  \includegraphics[width=\linewidth]{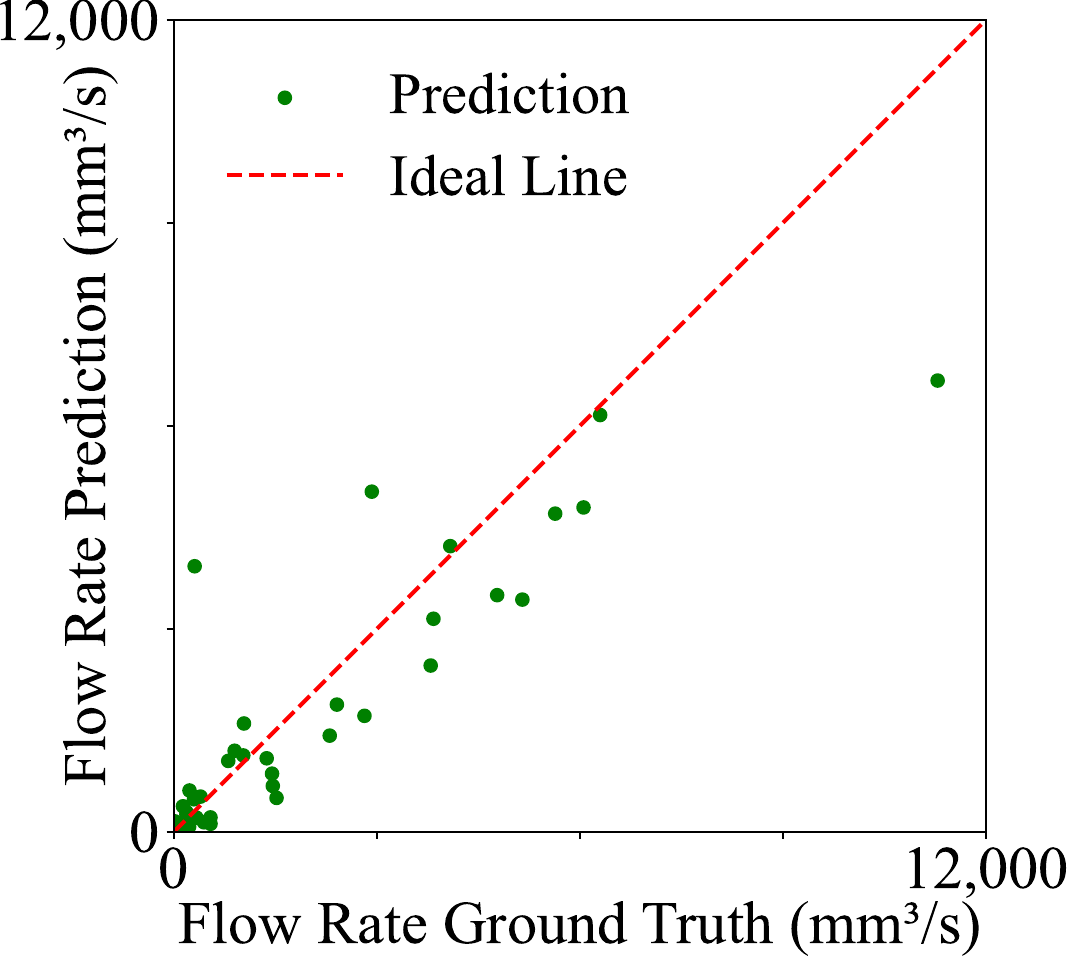}
  \centerline{(b) Flow rate}
\end{minipage}
\caption{Comparison of predicted and ground truth values for (a) pressure and (b) flow rate. The red dashed line represents the ideal regression line.}
\label{fig:scatterplot}
\end{figure}

\noindent  Performance comparison tables were visualized, and for evaluation, 100 nodes were randomly selected for visualization.

The model in this study achieved the following Pearson's correlation coefficients: 0.727 for pressure and 0.824 for flow rate, demonstrating superior performance compared to existing cerebral blood flow prediction models~\cite{ni2023voxel2hemodynamics}. Furthermore, the model achieved an accuracy of 84.8097 for pressure and 86.6405 for flow rate.

\begin{table}[ht]
\centering
\caption{Comparison of accuracy and Pearson's correlation coefficients for pressure and flow rate predictions using the GNN.}
\begin{tabular}{c c c}
        & \textbf{Pressure} & \textbf{Flow rate} \\ \hline
\textbf{Accuracy} & 84.8097      & 86.6405      \\ 
\textbf{Pearson's correlation coefficients} & 0.727 & 0.824 \\ 
\end{tabular}
\label{tab:performance_comparison}
\end{table}

\section{Discussion}
This paper applied a gROM to predict blood flow and pressure in human cerebral vasculature, utilizing MRA datasets from 35 individuals, augmented to 875 datasets. The study demonstrated the feasibility of adapting a model originally developed for cardiovascular and pulmonary vessels to cerebral arteries. While the gROM approach showed promise, it exhibited reduced accuracy when applied to previously unseen network configurations. The model particularly struggled with complex topologies due to stenosis in all datasets, underscoring the challenges in modeling pathological vascular networks. In conclusion, gROM shows significant potential for cerebral blood flow prediction but requires further refinement for complex vascular structures. Future work should focus on expanding datasets and improving augmentation techniques to enhance the model's generalizability and performance in pathological cases.

\section{Compliance with Ethical Standards}
This study was approved by the Institutional Review Board of Pusan National University Yangsan Hospital, with approval number 55-2024-044. Informed consent was obtained from all individual participants included in the study.
   
\section{Acknowledgments}
This work was supported by the Technology development Program(RS-2023-00303878) funded by the Ministry of SMEs and Startups(MSS, Korea). 

\bibliographystyle{IEEEbib}
\bibliography{refs}
\end{document}